\documentclass[aps,prl,twocolumn,superscriptaddress]{revtex4}
\usepackage{color}
\usepackage{amsmath,bm}
\usepackage{graphicx}

\begin{document}

\title{First-principles study of substitutional carbon pair and Stone-Wales defect complexes in boron nitride nanotubes}

\author{Gunn Kim}
\author{Jinwoo Park}
\author{Suklyun Hong}
\affiliation{Department of Physics and Graphene Research Institute, Sejong University, Seoul 143-747, Korea}
\date{\today}

\begin{abstract}
Using density functional theory, we study physical properties of boron nitride nanotubes (BNNTs) 
with the substitutional carbon pair defect. 
We also consider the Stone-Wales (SW) rearrangement of the C-C pair defect in the BNNT.
The formation energy of an SW defect of the carbon dimer is approximately 3.1 eV lower than that of the SW-transformed B-N pair in the
undoped BNNT. The activation energies show that the SW defect in the
C-doped BNNT may be experimentally observed with a higher
probability than in the undoped BNNT. Finally, we discuss the localized
states originating from the carbon pair impurities.
\end{abstract}

\maketitle

As a wide-band-gap nanomaterial\cite{Xu,Si,Blase}, hexagonal boron nitride (h-BN) is a stable crystalline form
consisting of equal numbers of boron and nitrogen atoms in a
honeycomb arrangement\cite{Watanabe,Rubio,Chopra}.
As it has considerably high thermal and chemical stabilities\cite{Golberg1}, it can be used for
high-temperature applications. The BN nanotube
(BNNT)\cite{Loiseau,Golberg2} is a nano-sized seamless cylinder that
can be considered a rolled-up h-BN sheet.
It also has a wide band gap that is almost independent of the tube diameter, and chirality\cite{Fuentes}.
For $sp^2$-bonded
carbon nanostructures such as fullerenes, carbon nanotubes and
graphene, the Stone-Wales (SW) transformation\cite{Stone} is believed  
to introduce topological defects or isomerization, which
results in a 90$^{\circ}$ rotation of two carbon atoms with respect to
the midpoint of the C-C bond.

When an SW defect is present in h-BN or BNNT, B-B and
N-N bonds should be created, and the formation of these bonds increases
the total energy of the system. Thus, the topological defect created by the
SW rearrangement is lacking. 
However, the formation of B-B and N-N
bond pairs can be avoided if there is a C-C pair defect in h-BN (or
BNNT). Using post-synthesis doping, carbon impurities can be substituted in the h-BN layers\cite{Krivanek, Wei1, Wei2}. 
During the formation of BNNTs and
h-BN sheets using the chemical vapor deposition (CVD), 
carbon atoms may act as substitutional defects, because
residual hydrocarbon may remain in the chamber, or carbon impurities
are solubilized in the metal catalyst. 
If carbon substitutional atoms congregate together in a local
area in h-BN (or a bundle of BNNTs) by segregation, they may form 
the interlayer (or intertube) conduction channels. As
mentioned above, the C-C pair defect in the h-BN network prohibits
the formation of B-B and N-N bonds during the SW transformation. 
As the SW-transformed defects may affect the mechanical and
electrical properties of the BN nanostrucutres, we should thus study these defects in the C-doped BN
sheets or BNNTs. This work may explain the formation of topological
defects in the carbon-doped BNNTs or h-BN layers that can be
produced in the CVD processes. In this paper, we present our
first-principles study of the structural and electronic properties of
the BNNTs containing two substitutional carbon defects.
The effect of carbon doping on the SW transformation is also examined.

\begin{figure}[b]
\centering
\includegraphics[width=8.0cm]{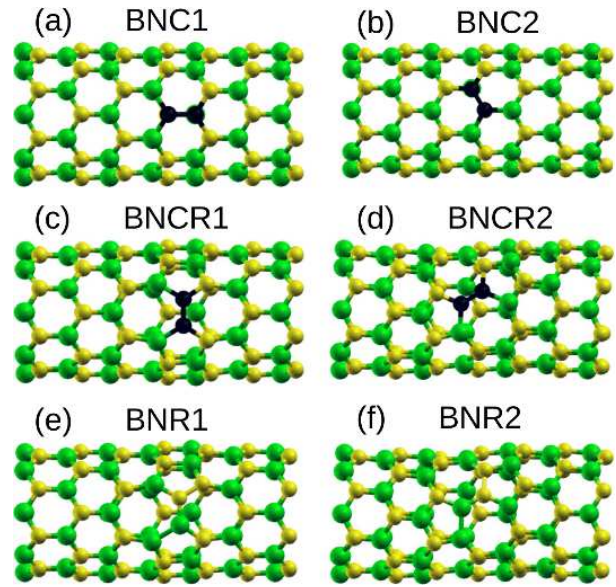}
\caption{Optimized structures of (8,0) BNNTs with a C-C pair substitutional defect or the SW defects.
(a) The BNNT with a C-C pair substitutional defect parallel to the tube axis, (b) the BNNT with a C-C pair substitutional defect with an oblique angle
with respect to the tube axis. (c) and (d) show the SW-transformed C-C pair defect from (a) and (b), respectively.
(e) and (f) are the BNNTs with SW defects which are parallel and tilted to the tube axis, respectively.
Yellow, green and black balls represent nitrogen, boron and carbon atoms, respectively.}
\label{fig1}
\end{figure}

We carried out total energy calculations for our model
systems based on the density functional theory\cite{Kohn}. The local
density approximation (LDA)\cite{Kohn-Sham} with spin polarization
was used for the exchange-correlation functional. 
Some calculations were repeated using the generalized gradient approximation.
We found that the exchange-correlation functional did not change our main conclusion 
that the SW defect in the C-doped BNNT may be observed with a higher probability than in the bare BNNT.
To study the energetics and the activation energy barrier in the SW
transformation of the systems, we used a plane wave basis set with an
energy cutoff of 400 eV and ultrasoft
pseudopotentials\cite{vanderbilt} implemented in the Vienna ab
initio simulation package (VASP)\cite{kresse1,kresse2}. The atomistic
models were relaxed until the residual forces on the atoms became smaller
than 0.03 eV/\AA. For the activation energy barrier, we employed the
nudged elastic band method\cite{Mills}. Eight replicas were
selected, including the initial and final configurations to
construct an elastic band. To plot the density of states (DOS) and
the wavefunctions with phase information, we used the OpenMX code\cite{Ozaki1}. 
In the calculations using the OpenMX code, the ionic
potentials were described with norm-conserving Troullier-Martins
pseudopotentials. Wavefunctions were expanded in a pseudoatomic
orbital basis set\cite{Ozaki2} with an energy cutoff of 120 Ry. As a
model system, we chose the (8,0) zigzag BNNT containing 48 B atoms
and 48 N atoms. The supercell size in the lateral direction was 25
\AA~ to avoid the interaction between the defects in neighboring
BNNTs, and the size in the axial direction was 12.96 \AA.

\begin{figure}[b]
\centering
\includegraphics[width=8.0cm]{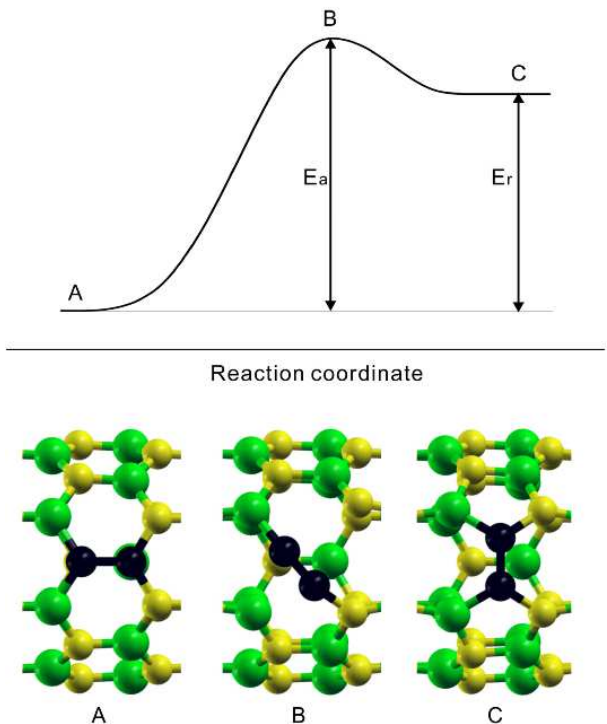}
\caption{(Top panel) Energy profile along the reaction coordinate in the SW transformation of BNNTs with a C-C pair substitutional defect or the SW defects.
The states ``A'', ``B'', and ``C'' represent the initial, transition, and final states in the SW transformations, respectively.
The activation and formation energies are denoted as $E_a$ and $E_r$, respectively.
(Bottom panel) The corresponding three states are shown for the SW transformation from BNC1 to BNCR1.
}
\label{fig2}
\end{figure}

\begin{table}[b]
\caption{Calculated activation ($E_a$) and formation ($E_r$) energies
for the SW transformation from the C-doped or the BNNTs to the BNNTs with SW defects.}
\begin{center}
\begin{tabular}{ccc}
\hline
               & $E_a$ (eV) & $E_r$ (eV)  \\
\hline
  BNC1 $\rightarrow$ BNCR1  & 5.50 & 2.51   \\
  BNC2 $\rightarrow$ BNCR2  & 5.64 & 2.17   \\
  BN1 $\rightarrow$ BNR1  & 7.04 & 5.61   \\
  BN2 $\rightarrow$ BNR2  & 7.63 & 5.72   \\
\hline

\end{tabular}
\end{center}

\end{table}



First, we optimized the C-doped and undoped BNNTs. The perfect BNNT
has a B-N bond length of $\sim$ 1.44 \AA. Our total energy
calculations show that the two substitutional carbon atoms tend to
bind to each other. This finding agrees with K. Yuge's report that the
lowest-formation-energy structures for monolayer
boron-carbon-nitride favor B-N and C-C bonds whereas they disfavor
B-C, C-N, B-B, and N-N bonds\cite{Yuge}. For C-doping, two types of
C-C pairs can be considered. As shown in Figure 1, one pair (BNC1) was 
parallel to the tube axis and the other (BNC2) had an oblique angle
with the tube axis. The bond length of the C-C pair was 1.37 \AA
~(1.39 \AA) for BNC1 (BNC2). Note that the bond length of benzene
(C$_6$H$_6$) has a very similar C-C bond length of 1.40 \AA. 
Although the model of BNC1 had the lowest energy among our model systems for the C-C pair, the energy difference between BNC1 and BNC2 was only $\sim$1 meV/atom. 
Thus, two structures may coexist.
In fact, by the chemical potential analysis, we found that the formation of carbon substitutional defects in the BNNTs, BNC1 or BNC2, is possible when the chemical potential of an isolated C-C pair is larger than by -18.02 eV or -17.90 eV. 
Note that the energy of an C-C pair is -10.09 eV.

In the C-C pair, the C atom bonded to two B atoms protruded out by
$\sim$ 0.1 \AA. The B-C bond length was 1.52 \AA~and the N-C bond
length was 1.46 \AA, which was due to the longer covalent radius of B
(r$_{covalent,B}$ = 0.84 \AA) than that of N (r$_{covalent,N}$ =
0.71 \AA). When the C-C bond was rotated by 90$^{\circ}$, the bond length slightly decreased to 1.35 \AA.
Here, the B-C bond length was increased to 1.61 \AA, and the N-C bond length
was slightly decreased to 1.44 \AA. After the SW transformation from
the BNC1 to the BNCR1 structure, the total energy of BNCR1 increased
by 2.5 eV, compared to that of BNC1. For this structure, the
activation energy from BNC1 to BNCR1 was 5.5 eV. Similarly, the
formation and activation energies were 2.2 eV and 5.6 eV, respectively, 
for the SW transformation from BNC2 to BNCR2 (see Table I).

For comparison, we also calculated the formation and activation
energies for the SW transformation in the undoped (8,0) BNNT as shown in Figure 2. 
For the SW transformation from BN1 to BNR1 ({\textit or} from BN2 to BNR2),
the formation energy was 5.6 eV (5.7 eV), and the activation energy was
7.0 eV (7.6 eV). Here, a B-B bond and a N-N bond are produced, and
their lengths are 1.73 \AA~(B-B) and 1.47 \AA~(N-N), respectively. 
We found that the rotating C-C bond (1.24 \AA) was shorter than the rotating B-N bond (1.29 \AA) in the transition states.
This implies that the shorter bond results in a lower activation energy barrier during the SW transformation in a small area. 
The high formation energy is due to the unstable B-B and N-N bonds as mentioned
earlier. Interestingly, electrons were accumulated in the N atom of
the rotating B-N pair in the transition state for the undoped BNNT. In addition, our charge analysis
revealed that electron donation from the C-C pair defect in the
C-doped BNNT slightly weakens the dangling-bond-like character of
the B and N atoms in the transition configuration. In our previous
study of the SW transformation in M@C$_{60}$ (M = K, Ca, and La), we
similarly observed that the electron donation by the
incorporated metal atom lowers the activation energy
barrier\cite{WChoi}. These findings can explain the lower activation energy barrier of the C-C pair defect compared to that
of the B-N pair. Therefore, when the C-doped BNNT {\em does} exist, e.g., from the sample
preparation, we can conclude that the probability that the SW defect
is observed is higher in the C-doped BNNT than in the undoped BNNT.

\begin{figure}[t]
\centering
\includegraphics[width=8.0cm]{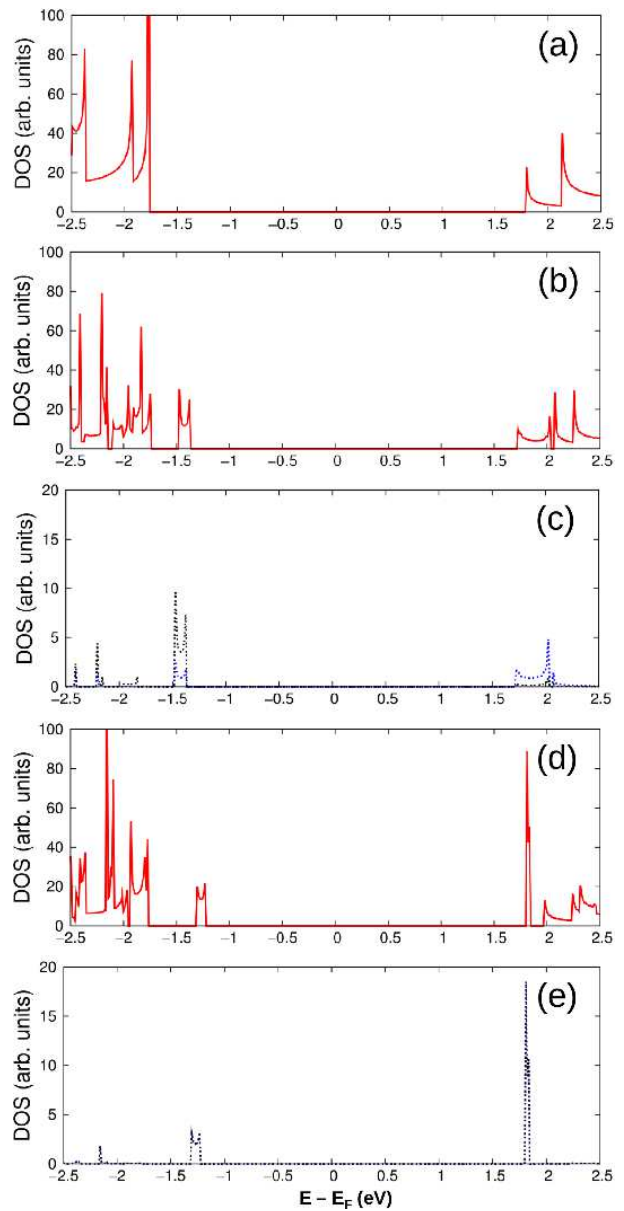}
\caption{(a) Total DOS of the perfect (8,0) BNNT.
Total DOS of BNNTs with (b) the C-C pair substitutional defect or (d) the SW defects.
Dotted lines in (c) and (e) indicate the projected DOS of two C atoms in the C-C pair defect in each case.
}
\label{fig3}
\end{figure}

To check a charge doping effect, we calculated doping of a hole (charge +$e$ state) for all the model structures 
(BNR1, BNR2, BNCR1, and BNCR2), and an electron doping case (charge -$e$ state) for BNR2 structure. 
When a hole was doped, the activation energy barriers were {\it reduced} from 7.04 eV to 6.91 eV for BNR1, 7.63 eV to 7.18 eV for BNR2, 
5.50 eV to 5.23 eV for BNCR1, and 5.64 eV to 5.52 eV for BNCR2, respectively. 
For an electron doping in BNR2, and the activation energy was {\it decreased} from 7.63 eV to 6.69 eV. 
Therefore, we can conclude that charge doping affects the activation energy barriers of the SW transformation.

In a growth process of the BNNTs, carbon atoms may be incorporated as residual impurities. 
Or one can produce carbon substitutional defects in the BNNTs on purpose for doping. 
Our results demonstrate that two carbon atoms far from each other tend to become closer and form a C-C pair. 
The energy difference between the C-C pair and two separate carbon atoms was larger than 5 eV in our LDA calculation. 
On the other hand, some metal atoms or molecules such as O$_2$ and Br$_2$ may be adsorbed in the vicinity of the C-C pair. 
Then, electron transfer takes place between the adsorbate and the nanotube. 
As mentioned above, charge doping enables the activation energy barrier to become lowered and the probability of the SW transformation to become higher.

\begin{figure}[b]
\centering
\includegraphics[width=8.0cm]{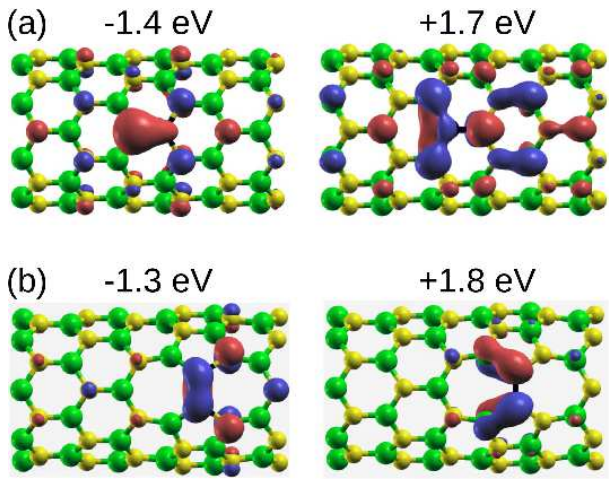}
\caption{Wavefunctions originating from the defect states in the BNNTs with (a) the substitutional C-C pair defect (BNC1)
and (b) the SW-transformed C-C pair defect (BNCR1).
Red and blue colors represent the opposite signs in the wavefunctions at the $\Gamma$-point.
}
\label{fig4}
\end{figure}

According to our LDA calculation, the energy band gap of the perfect 
(8,0) BNNT is 3.6 eV as shown in Figure 3a. R. Wu et al.
reported that the BNNT could have a magnetic moment due to the carbon
doping\cite{Wu}. They considered only one carbon atom in the
supercell. However, our calculational results demonstrated that 
the magnetic moment does not occur in the BNNT with the C-C pair defect.
According to a recent report by N. Berseneva et al., the
magnetic moment is zero when the numbers of carbon atoms replaced by
nitrogen and boron atoms are
identical\cite{Berseneva}. Figure \ref{fig3}b shows the total DOS of BNNTs with the C-C pair substitutional
defect (e.g., BNC1), which clearly demonstrates that a few localized
states occur in the band gap in the presence of the C-C pair
defect. As shown in Figure~\ref{fig3}c, the projected DOS of the two
C atoms in the C-C pair defect indicates that the gap states
originate from the substitutional carbon pair defects. When a C-C
pair forms a SW defect by a 90$^{\circ}$ rotation as seen in BNCR1, the
electronic states originating from the SW defect have flatter bands
with more localized characters. These characteristics are reflected in the sharp and
narrow peaks in the DOS in the gap (Figure~\ref{fig3}d and \ref{fig3}e).

Figure~\ref{fig4} shows wavefunction characters resulting from the
defect states in the BNNTs with the substitutional carbon pair
defect and its SW-transformed defect. The occupied state at $-$1.4 eV
and the unoccupied state at $+$1.7 eV shown in Figure \ref{fig4}a
are the localized defect states originating primarily from the C-C pair of the BNC1
structure. The filled state has a bonding character, whereas the
unfilled state has a antibonding character at the C-C pair. 
As displayed in Figure \ref{fig4}a, the electron density at the C atom bonded to B atoms is somewhat
higher than that at the C atom bonded to N atoms at $-$1.4 eV.
Additionally, the B-C bonds are in the bonding states but the C-N bonds
are in the antibonding states. On the other hand,
Figure~\ref{fig4}b shows the localized defect states at $-$1.3 eV
and $+$1.8 eV from the SW-transformed C-C pair defect of the BNCR1
structure. Similar to the BNC1 case, bonding and antibonding
characters of the carbon pair appear at $-$1.3 eV and $+$1.8 eV,
respectively. Note that there is no mirror (reflection) symmentry for
the BNNT with respect to the plane containing the midpoint of the
C-C bond. However, the mirror symmetry is manifest with respect to
the planes containing the C-C bond itself (as in BNC1) as well as
the midpoint of the SW-transformed C-C bond (as in BNCR1). These
features were retained for two localized states as shown in
Figure~\ref{fig4}a and also for a localized state at $-$1.3 eV as shown in
Figure~\ref{fig4}b.

In summary, we have introduced substitutional carbon pair defects in
BNNTs to examine the modifications in the structural and
electronic properties of the nanotube. Our results demonstrate that
the two carbon atoms prefer to bind to each other with a bond length
of $\sim$1.37 \AA. For the SW transformation of the C-C pair defect,
the formation energy of the SW defect is approximately 2.5 eV and the
activation energy is approximately 5.5 eV. Both the formation and
activation energies of the C-C bond are much lower than those of
the B-N bond in the undoped BNNT in the SW transformation, which are $\sim$5.5 eV and
$\sim$7.0 eV, respectively. Consequently, it would be very difficult
to find a topological defect such as the SW defect in the perfect 
BNNT. Conversely, such a topological defect could be observed with a
higher probability in the C-rich defective BNNTs.

\section*{Acknowledgments}
This research was supported by the Priority Research Center Program
(2011-0018395), the Converging Research Center Program
(2011K000620), and the Basic Science Research Program
(KRF-2008-313-C00217) through the National Research Foundation of
Korea (NRF) funded by the Ministry of Education, Science and
Technology(MEST). G. K. acknowledges support by the Basic Science
Research Program through MEST/NRF (2010-0007805).


\begin{thebibliography}{99}

\bibitem{Xu}
Y.N. Xu, W.Y. Ching, Phys. Rev. B 44 (1991) 7787. 

\bibitem{Si}
M.S. Si, D.S. Xue, Phys. Rev. B 75 (2007) 193409. 

\bibitem{Blase}
X. Blase, A. Rubio, S.G. Louie, M.L. Cohen, Europhys. Lett. 28 (1994) 335.

\bibitem{Watanabe}
K. Watanabe, T. Taniguchi, H. Kanda, Nature Mater. 3 (2004) 404.

\bibitem{Rubio}
A. Rubio, J.L. Corkill, M. L. Cohen, Phys. Rev. B 49 (1994) 5081.

\bibitem{Chopra}
N.G. Chopra, R.J. Luyken, K. Cherrey, V.H. Crespi, M.L. Cohen, S. G. Louiea, A. Zettl, Science 269 (1995) 966.

\bibitem{Golberg1}
D. Golberg, Y. Bando, C.C. Tang, C.Y. Zhi, Adv. Mater. 19 (2007) 2413.

\bibitem{Loiseau}
A. Loiseau, F. Willaime, N. Demoncy, G. Hug, H. Pascard, Phys. Rev. Lett. 76 (1996) 4737.


\bibitem{Golberg2}
D. Golberg, Y. Bando, W. Han, K. Kurashima, T. Sato, Chem. Phys. Lett. 308 (1999) 337.

\bibitem{Fuentes}
G.G. Fuentes, E. Borowiak-Palen, T. Pichler, X. Liu, A. Graff, G. Behr, R.J. Kalenczuk, M. Knupfer. J. Fink, Phys. Rev. B 67 (2003) 035429.

\bibitem{Stone}
A.J. Stone, D.J. Wales, Chem. Phys. Lett. 128 (1986) 501.


\bibitem{Krivanek}
O.L. Krivanek et al., Nature (London) 464 (2010) 571.

\bibitem{Wei1}
X. Wei, M. Wang, Y. Bando, D. Golberg, J. Am. Chem. Soc. 132 (2010) 13592.

\bibitem{Wei2}
X. Wei, M. Wang, Y. Bando, D. Golberg, ACS Nano 5 (2011) 2916.

\bibitem{Kohn}
P. Hohenberg, W. Kohn, Phys. Rev. 136 (1964) B864.

\bibitem{Kohn-Sham}
W. Kohn, L.J. Sham, Phys. Rev. 140 (1965) A1133.

\bibitem{vanderbilt}
D. Vanderbilt, Phys. Rev. B 41 (1990) 7892.

\bibitem{kresse1}
G. Kresse, J. Hafner, Phys. Rev. B 47 (1993) R558.

\bibitem{kresse2}
G. Kresse, J. Furthm\"{u}ller, Phys. Rev. B 54 (1996) 11169.


\bibitem{Mills}
G. Mills, H. J\'onsson, Phys. Rev. Lett. 72 (1994) 1124.


\bibitem{Ozaki1}
T. Ozaki, Phys. Rev. B. 64 (2001) 195110.


\bibitem{Ozaki2}
T. Ozaki, Phys. Rev. B. 67 (2003) 155108.

\bibitem{Yuge}
K. Yuge, Phys. Rev. B 79 (2009) 144109.

\bibitem{WChoi}
W.I. Choi, G. Kim, S. Han, J. Ihm, Phys. Rev. B 73 (2006) 113406.

\bibitem{Wu}
R.Q. Wu, L. Liu, G.W. Peng, Y.P. Feng, Appl. Phys. Lett. 86 (2005) 122510.

\bibitem{Berseneva}
N. Berseneva, A.V. Krasheninnikov, R.M. Nieminen, Phys. Rev. Lett. 107 (2011) 035501.

\end{thebibliography}
\end{document}